\documentclass[pra,preprint,fleqn]{revtex4-1}
\usepackage{CJK}
\usepackage{graphicx}
\usepackage{amsmath}
\usepackage{amssymb}
\usepackage{amsmath}
\usepackage{color}
\usepackage{bm}
\usepackage{latexsym}
\usepackage{hyperref}
\usepackage{graphicx}
\usepackage{epstopdf}
\usepackage{subfigure}
\usepackage[normalem]{ulem}
\usepackage{tikz}
\usetikzlibrary{positioning}
\newcommand{\void}[1]{}

\newcommand{\non}{\nonumber}

\newcommand{\be}{\begin{equation}}
\newcommand{\ee}{\end{equation}}
\newcommand{\ket}[1]{\ensuremath{\left|{#1}\right\rangle}}

\begin{document}
\title{Exact variational dynamics of the multi-mode Bose-Hubbard model based
on SU($M$) coherent states}
\author{Yulong Qiao}
\affiliation{Max-Planck-Institut f\"ur Physik Komplexer
Systeme, N\"othnitzer Str. 38, D-01187 Dresden, Germany}
\affiliation{Institut f\"ur Theoretische Physik, Technische Universit\"at
Dresden, D-01062 Dresden, Germany}
\author{Frank Grossmann}
\affiliation{Institut f\"ur Theoretische Physik, Technische Universit\"at
Dresden, D-01062 Dresden, Germany}
\date{\today}

\begin{abstract}
We propose a variational approach to the dynamics of the
Bose-Hubbard model beyond the mean field approximation.
To develop a numerical scheme, we use a discrete overcomplete
set of Glauber coherent states and its connection to the generalized coherent
states studied in depth by Perelomov [A. Perelomov, {\it  Generalized Coherent States and Their Applications}, Springer-Verlag (Berlin, 1986)].
The variational equations of motion of the generalized coherent state parameters as well as of the coefficients in an expansion of the wavefunction
in terms of those states are derived and solved for many-particle problems
with large particle numbers $S$ and increasing mode number $M$. For $M=6$ it is revealed that
the number of complex-valued parameters that have to be propagated is more
than one order of magnitude less than in an expansion in terms of Fock states.
\end{abstract}

\maketitle

\section{Introduction}

The mathematical foundation of the phase space formulation of physical systems
with Lie group symmetries has been considerably widened by the works of Brif and Mann \cite{BM98,BM99}.
Based on their progress, the description of the Bose-Hubbard (BH) dynamics in terms of phase space distributions has received new impetus from the works of Korsch and collaborators
\cite{TWK08,TWK09}. Therein,  equations of motion for the $P$- as well as the $Q$-function of quantum optics \cite{ZFG90,GaZo} have been derived and solved for small site numbers $M$. The basis functions that were found appropriate for
the treatment of the particle number conserving dynamics are so-called SU($M$) coherent states (CS), also referred to as generalized coherent state (GCS) \cite{Pere}. These have been investigated and favored in the same context by Buonsante and Penna in their review \cite{BP08},
whose focus is on variational mean field methods.
A more recent review with a focus on SU(2) CS, introduced as atomic coherent states in \cite{ACGT72}, is given in \cite{KRG17}. An extension of the formalism towards dissipative Lindblad type equations in terms of $P$ functions has been given in \cite{MLLS20}.

In parallel, the use of discrete, complete von Neumann type sets of the more
``standard'' Glauber coherent states, whose position representation are Gaussian wavepackets, has been applied to a flurry of different physical situations, ranging from electron dynamics in atoms to nuclear dynamics in molecules as well as to nonadiabatic (combined electron-nuclear motion), as
reviewed in \cite{camp99,Mi01,TW04,Kay05,Richings2015,irpc21}.
The equations of motion of the coherent state parameters as well as of the coefficients in the
expansion of the wavefunction in terms of those states are usually derived from a variational principle and possibly undergo additional approximations.
It has turned out in numerical investigations that the
use of a surprisingly small number of CS basis functions leads to converged results, e.g., in spin-boson type problems tackled by the so-called multi Davydov D2 Ansatz \cite{Wang2017,prb20,pra20},
\textcolor{black}{or in the method called coupled coherent states for indistinguishable bosons \cite{GrSh19}.}

Furthermore, in \cite{LF14}, a generalization of the multi-configuration time-dependent Hartree
method for bosons (MCTDHB) \cite{ASC08} based on McLachlan's variational principle has been given.
The time-dependent permanents used there are based on orthogonal orbitals, however.
Previous approaches that have used Glauber coherent state based semiclassical propagators
for BH dynamics have been restricted to small mode numbers \cite{SiSt14,Retal16},
as is also the case for
semiclassical approaches based on SU($M$) CS \cite{ViAg11,KCV13}, whereas mean field
approaches as the ones discussed in \cite{BP08} rely just on a single basis function (i.e. trivial multiplicity).

It has been worked out in \cite{RMW10,RMCKW10}, that Gaussian CS are promising basis functions also for full fledged dynamical calculations for BH dynamics beyond the semiclassical
propagation or the Gross-Pitaevskii level \cite{PSG02}. \textcolor{black}{Similarly, imaginary time propagation to find the ground state as well as real-time propagation to extract the excitation spectrum using Gaussians has been
performed recently for BH systems \cite{Getal19}. In addition, in the two mode case, the use of a single time-dependent atomic coherent state has been promoted in \cite{WMBS21}.}
These successes as well as that of the CS basis functions alluded to above, leads us to investigate the question if also for the fixed particle number {\it generalized} coherent states
multi-mode BH dynamics can be treated by a numerically exact variational approach, based on a multitude of those GCS.

The presentation is organized as follows: First, in Section II, we review the one-dimensional
BH model and the relation between the Glauber (field) coherent states \cite{Loui}
and the (generalized) SU($M$) coherent states \cite{Pere}. This latter review is necessary
in order to get a handle on the discretization of the representation of unity in terms of these
states. In Section III, we then derive the variational equations of
motion for the GCS parameters using the Lagrangian of the time-dependent variational principle.
As a proof of principle in Sect. IV we finally solve the dynamical
problem for several realizations of the BH model with different (relatively large)
mode number $M$ and particle number $S> M$ and compare to results gained from an expansion in
Fock states where this is still feasible. Conclusions are given in Sect. V, while in the appendix
some properties of the GCS that are needed in the main text, as well as the matrix form of the variational equations
\textcolor{black}{and a detailed convergence study} are gathered.

\section{Bose-Hubbard model and completeness of SU($M$) coherent states}

In the following we are constructing a discrete set of GCS
for use in a variational approach to the dynamics of the BH model, which is going to
be reviewed first.

\subsection{Bose-Hubbard model in one dimension}

In this presentation, we are aiming at a treatment in terms of the variational principle
of the dynamics of the one-dimensional linear chain BH model \cite{JZ05}
\begin{align}
\label{eq:BH}
 \hat{H} =-J\sum_{j=1}^{M-1}\left[(\hat a_j^\dag \hat a_{j+1}+h.c.)\right]
    +\frac{U}{2}\sum_{j=1}^M \hat a_j^{\dag 2}\hat a_j^{2}
    +\frac{K}{2}\sum_{j=1}^M(j-j_0)^2\hat a_j^\dag \hat a_j,
\end{align}
written here in normal-ordered form in terms of bosonic creation and annihilation operators,
$\hat{a}_j^\dagger$ and $\hat{a}_j$, which fulfill the
commutation relation $[\hat{a}_i,\hat{a}_j^\dagger]=\hat \delta_{ij}$ and their action on
number states $\{|n_j\rangle\}, n_j=0,1,\dots$ is given by
\be
\label{eq:creann}
\hat{a}_j|n_j\rangle=\sqrt{n_j}|n_j-1\rangle,
\\
\hat{a}_j^\dagger|n_j\rangle=\sqrt{n_j+1}|n_j+1\rangle.
\ee
The hopping matrix element $J$
is determined by the tunneling probability between nearest neighbors, whereas $U$ denotes the on-site interaction, and $K$ is due to an external, harmonic trapping potential (centered around $j_0$),
which will, however, be set to zero for all the numerical results to be presented below.
The above Hamiltonian is a widely studied model for the dynamics of spinless particles in optical lattices \cite{BDZ08}. An experimental realization of a quantum phase transition has been reached by varying the model's parameters \cite{Getal02}.
\textcolor{black}{Furthermore, Kolovsky has given a review of the spectral properties of the BH Hamiltonian in the light of quantum chaology \cite{Kolo16}.}

In \cite{TWK08} it has been argued that an analysis in terms of flat space and the use of the
corresponding Glauber coherent states to describe particle number preserving BH dynamics is inadequate. The reason given, in the case of $M$ sites (or modes), is that the dynamical group is spanned by the normally ordered operators
\be
\label{eq:no}
\hat{a}_i^\dagger\hat{a}_j,\qquad i,j\in\{1,2,\dots M\}
\ee
and is equivalent to the special unitary group SU($M$). Therefore the corresponding GCS should be employed to investigate the dynamics. We note in passing that,
in terms of the normally ordered operators from Eq.\ (\ref{eq:no}), the operators
$\hat a_j^{\dag 2}\hat a_j^{2}$ from Eq.\ (\ref{eq:BH}) are written as
$\hat n_j(\hat n_j-1)$, with the number operators $\hat{n}_j=\hat{a}_j^\dag\hat{a}_j$.

\subsection{Discrete grid of generalized coherent states}
\label{ssec:disc}

An insightful didactic discussion of the group theoretical construction of coherent states is
given in \cite{ZFG90}. Starting by revisiting the simple case of Glauber field coherent states as the
coherent states of the Heisenberg Weyl group, more complex situations are discussed, where the
group theoretical approach will be more advantageous for the understanding of the topological structure of the corresponding coherent states than in the simple case.
An open question remains as to the completeness of a discrete set of coherent states in the general case, however.

The Glauber coherent states $\ket{z}$ (single site case) can be defined via the eigenvalue equation
\begin{align}
\hat{a}\ket{z}&=z\ket{z}
\end{align}
with the complex eigenvalue
\begin{align}
z&=\frac{\gamma^{1/2}q+{\rm i}\gamma^{-1/2}p}{\sqrt{2}}.
\end{align}
in terms of position and momentum (phase space) and width parameter $\gamma$.
These states consist of a Poissonian superposition of number states \cite{Loui}.

Exactly fifty years ago, two independent contributions have proven the statement that a discretized version
of the unit operator is given by
\be
\label{eq:unity_1}
\sum_{k,l}|z_k\rangle({\mathbf \Omega}^{-1})_{kl}\langle z_l|=\hat{I},
\ee
with the overlap matrix ${\mathbf \Omega}$ with elements
\be
\Omega_{kl}=\langle z_k|z_l\rangle,
\ee
if the requirement
\be
z_k=\beta(m+{\rm i}n),
\ee
with $k=(m,n)$  where $m,n$ are integers and $0<\beta\leq \sqrt{\pi}$, for the spacing of the grid points is met \cite{Pere71,BBGK71}. Physically, this means that the cells in $(p,q)$
phase space that are spanned by the grid points must have an area less than or equal to
the Planck cell area of $2\pi$ in the present units (where $\hbar=1$). The limiting case of
$\sqrt\pi$ for the spacing of the grid points has already been postulated long before by
von Neumann \cite{vonNeu,BoZa78}.

As mentioned above, for the BH model, generalized SU($M$) coherent states are
the appropriate coherent states, staying within the particle number conserving subspace of
the dynamics. In order to get a handle on their completeness,
we use the expansion of a Glauber coherent state in terms of the generalized CS.
The representation of the multimode GCS that is most appropriate to this
end is \cite{BP08}
\begin{align}
\label{wh}
|S,\vec{\xi}\rangle=\frac{1}{\sqrt{S!}}\Big(\sum_{i=1}^M\xi_ia_i^\dag\Big)^S|0,0,\cdots,0\rangle,
\end{align}
where $|0,0,\cdots,0\rangle$ denotes the multi-mode vacuum state and $S$ is the number of bosons of the GCS. Some properties of the GCS that will be needed later-on are
gathered in Appendix \ref{app:prop}.
The set of complex numbers $\{\xi_i\}$ are characteristic parameters of the GCS, and they satisfy the normalization condition
\begin{align}
\label{eq:norm}
\sum^{M}_{i=1}|\xi_i|^2=1,
\end{align}
where $M$ represents the number of different modes present in the BH model of Eq.\ (\ref{eq:BH}). Together with the total number of bosons $S$ (\textcolor{black}{not restricted to, but} herein chosen to be larger than $M)$, this
determines the dimension of the Hilbert space
\begin{align}
\label{eq:choose}
C^{F}= {M+S-1 \choose M-1}= {M+S-1 \choose S}=\frac{(M+S-1)!}{S!(M-1)!}
\end{align}
spanned by the Fock states \cite{Kolo16}. The relation of the above definition of the GCS to the
group theoretical formulation for $M>2$ is obscured by the necessity to disentangle exponentiated
operators \cite{BP08}. It is helpful to write out the definition in Eq.\ (\ref{wh}) explicitly for small numbers $M=2<S$
to convince oneself of the fact that the GCS is a superposition of Fock states with coefficients to be determined from a generalized binomial formula.

As shown in \cite{BP08}, and as can be proven by Taylor expansion of the exponential function,
the multimode Glauber coherent state $|Z\rangle=\prod_{i=1}^M|z_i\rangle$,
and the GCS are related by
\begin{align}
|Z\rangle\non\ &=e^{-\frac{1}{2}\sum_{i=1}^M|z_i|^2}e^{\sum_{i=1}^Mz_ia_i^\dag}|0,0,\cdots,0\rangle\\
\non &=e^{-\frac{1}{2}\sum_{i=1}^M|z_i|^2}\sum_{S=0}^{\infty}\frac{1}{S!}(\sum_{i=1}^Mz_ia_i^\dag)^S|0,0,\cdots,0\rangle\\
\non\ &=e^{-\frac{\tilde{N}}{2}}\sum_{S=0}^{\infty}\frac{\tilde{N}^{\frac{S}{2}}}{\sqrt{S!}}\frac{1}{\sqrt{S!}}\Big(\sum_{i=1}^M{\frac{z_i}{\sqrt{\tilde{N}}}}a_i^\dag\Big)^S|0,0,\cdots,0\rangle\\
&=e^{-\frac{\tilde{N}}{2}}\sum_{S=0}^{\infty}\frac{\tilde{N}^{\frac{S}{2}}}{\sqrt{S!}}|S,\vec{\xi}\rangle.
\end{align}
Here, $\tilde{N}=\sum_{i=1}^M|z_i|^2$ denotes the average number of bosons in $|Z\rangle$. We note that the relationship between $\xi_i$ and $z_i$ is $\xi_i=\frac{z_i}{\sqrt{\tilde{N}}}$.
Thus it is natural to construct a one-to-one map between the sets $\{\xi_{k,i}\}$ and $\{z_{k,i}\}$ (here the first index denotes the basis function discretization index and  second index denotes the mode index)
to be used in a discretized version of the unit operator.

To this end, we first generalize the completeness relation for the single site case from Eq.\ (\ref{eq:unity_1}) for multimode Glauber coherent states:
If the complex grid $L_{P,Q}^{M}$ which lies in the multidimensional complex $Z$-plane satisfies the multidimensional Planck cell condition, the multimode coherent states will form an overcomplete set and they obey the closure relation
\begin{align}
 \sum_{k,l}A_{k,l}|Z_k\rangle\langle Z_l|=\hat I, \indent Z_k, Z_l\in L_{P,Q}^{M},
\end{align}
where
\begin{align}
\label{eq:overlap}
\bm{A}=\bm{\Omega}^{-1}, \Omega_{kl}=\langle Z_k|Z_l\rangle,
\end{align}
with the multi-dimensional direct product CS $|Z_k\rangle=|z_{k1},z_{k2},\cdots z_{kM}\rangle$.
A heuristic proof of the closure relation can be obtained in the following way: Given any two coherent states $|Z_m\rangle$ and $|Z_n\rangle$ and inserting
unity in the multi degree of freedom case yields
\begin{align}
\langle Z_m|Z_n\rangle &\non =\sum_{k,l=1}^NA_{kl}\langle Z_m|Z_k\rangle\langle Z_l|Z_n\rangle\\
&=\sum_{k,l=1}^N\langle Z_m|Z_k\rangle A_{kl}\langle Z_l|Z_n\rangle
\end{align}
This equation has the matrix form $\bm{\Omega}=\bm{\Omega}\bm{A}\bm{\Omega}$ with the overlap matrix ${\bf \Omega}$ from Eq.\ (\ref{eq:overlap}) above.
With the definition of the matrix ${\bf A}$ as the inverse of the overlap matrix, the matrix equation becomes an identity.

From this, we infer that if a set of states is complete in the whole Hilbert space,
these states are also complete for any subspaces which belong to the Hilbert space. In our present case, we take the subspace to be the one consisting of multimode Fock states
$\{|n_S\rangle=|n_1,n_2,\cdots\rangle|n_1,n_2,\cdots\in\mathcal{N}, \langle n_S|\hat{N}|n_S\rangle=S,\hat{N}=\sum_i\hat n_i\}$ with fixed number of bosons $S$.

Since the Glauber coherent state can be written as a sum of SU($M$) coherent states, according to
\begin{align}\label{sumexpress}
|z_{k1},z_{k2},\cdots,z_{kj},\cdots\rangle=e^{-\frac{\tilde{N}}{2}}\sum_{S=0}^{\infty}\frac{\tilde{N}^{\frac{S}{2}}}{\sqrt{S!}}|S,\xi_{k1},\xi_{k2},\cdots,\xi_{kj},\cdots\rangle,
\end{align}
there is a map $z_{kj}\rightarrow \xi_{kj}=\frac{z_{kj}}{\sqrt{\sum_{i=1}^M|z_{ki}|^2}}$. Unlike the $|z_{k1},z_{k2},\cdots,z_{kj},\cdots\rangle$ where the parameters $z_{k1},z_{k2},\cdots,z_{kj},\cdots$ are mutually independent,
the $\xi_{k1},\xi_{k2},\cdots,\xi_{kj},\cdots$ are correlated due to the normalization factor $\frac{1}{\sqrt{\sum_{i=1}^M|z_{ki}|^2}}$, however.\\
It is obvious that if the $z_{k1},z_{k2},\cdots,z_{kj},\cdots$ form a complex grid fulfilling the completeness criterion, then the corresponding $\xi_{k1},\xi_{k2},\cdots,\xi_{kj},\cdots$ will also form a
complete basis
\begin{align}
\Big\{|S,\xi_{k1},\xi_{k2},\cdots,\xi_{kj},\cdots\rangle|\xi_{kj}=\frac{z_{kj}}{\sqrt{\sum_{i=1}^M|z_{ki}|^2}}, z_{kj}\in L_{P,Q}^{M}\Big\}.
\end{align}
for the subspace $\{|n_S\rangle=|n_1,n_2,\cdots\rangle|n_1,n_2,\cdots\in\mathcal{N}, \langle n_S|\hat{N}|n_S\rangle=S,\hat{N}=\sum_i\hat n_i\}$.

\begin{figure}[h]
  \centering
  \includegraphics[width=3.4in]{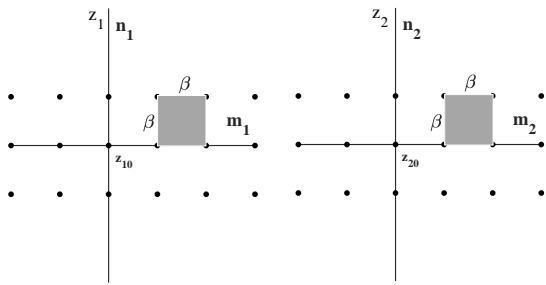}\\
  \caption{The complex grids for $z_1$ and $z_2$ in the case of two modes. They are centered around $\{z_{01},z_{02}\}=\{\xi_{01},\xi_{02}\}$ given by the GCS parameters of the initial state.}
  \label{compg}
\end{figure}

\begin{figure}[h]
  \centering
  \includegraphics[width=3.4in]{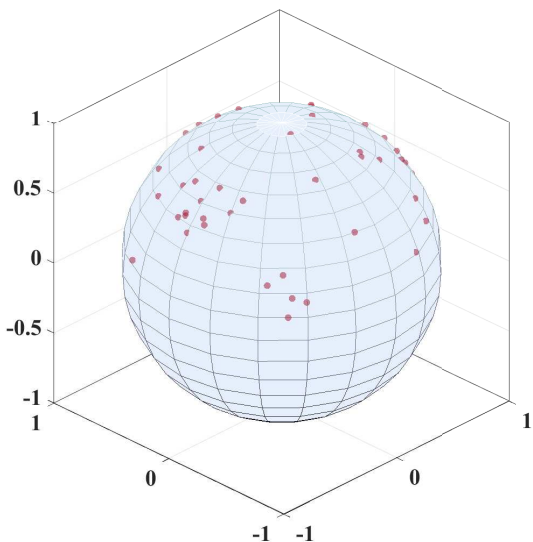}\\
  \caption{Some of the 50 grid points $\{\xi_{k1},\xi_{k2}\}$ on the surface of the unit sphere. There are more points on the back, which are invisible from the present perspective.}\label{comps}
\end{figure}

We now consider the two-mode case as an instructive example. The two initially independent complex grids for $z_{k1}$ and $z_{k2}$ are shown in Fig.~{\ref{compg}}. Taking samples from these two grids for $z_{k1}$ and $z_{k2}$ randomly at the same time, every produced pair of $\{z_{k1},z_{k2}\}$ forms a two-mode Glauber coherent state
$|z_{k1},z_{k2}\rangle$. The $\{z_{k1},z_{k2}\}$ correspond to a set of couples of parameters $\{\xi_{k1},\xi_{k2}\}=\{\frac{z_{k1}}{\sqrt{|z_{k1}|^2+|z_{k2}|^2}},\frac{z_{k2}}{\sqrt{|z_{k1}|^2+|z_{k2}|^2}}\}$ for the SU(2) coherent states, but due to the mutual correlation between $\xi_{k1}$ and $\xi_{k2}$,
they cannot be presented on two independent complex planes. We are now rewriting the GCS parameters in the form
\begin{align}
\xi_{k1}&=\cos\left(\frac{\theta_{k}}{2}\right)
\\
\xi_{k2}&=\sin\left(\frac{\theta_{k}}{2}\right)e^{i\phi_k},
\end{align}
where $0\leq\theta_k\leq\pi$ and $0\leq\phi_k\leq2\pi$ are the angles and the relative phases, respectively. \textcolor{black}{Because we have just two angles for every $k$ and because of the factor of 1/2 that comes along with the angle $\theta_k$, this allows us to represent
the GCS parameters as points on the northern hemisphere of the Bloch sphere.}
\textcolor{black}{For a visualization, we first generated 50 random pairs of $\{z_{k1},z_{k2}\}$.
Using the angles  $\{\theta_k,\phi_k\}$, the corresponding 50 pairs of $\{\xi_{k1},\xi_{k2}\}$
can be expressed via $\{\cos(\theta_k/2)\cos(\phi_k),\cos(\theta_k/2)\sin(\phi_k),\sin(\theta_k/2)\}$, as displayed in Fig.~(\ref{comps}).}

The procedure just presented can be generalized to more than just two modes and will allow us to perform numerical calculations with a finite, presumably
small number of GCS for the BH model. Before we can do so, we have to derive the relevant
equations of motion, however.

\section{Lagrangian formulation of the variational dynamics}

Generalizing the variational coherent state Ansatz reviewed in \cite{irpc21} by using the GCS discussed above, we are led to the expression
\begin{align}
 |\Psi(t)\rangle=\sum_{k=1}^NA_k(t)|S,\vec{\xi}_k(t)\rangle
 \label{eq:var}
\end{align}
for the wavefunction, where $|S,\vec{\xi}_k(t)\rangle=|S,\xi_{k1}(t),\xi_{k2}(t),\cdots\xi_{kM}(t)\rangle$ and $M$ is the
number of modes in the BH Hamiltonian of Eq.\ (\ref{eq:BH}). The initial set of basis functions $\{|S,\vec{\xi}_k(0)\rangle\}$
is the  one constructed in the previous section.
Both the expansion coefficients $\{A_k(t)\}$ as well as the basis functions are time-dependent.
The multiplicity parameter $N$ determines the number of interacting, complex parameters whose dynamics is to be determined.
It is given by $(M+1)N$ \textcolor{black}{and will have to be compared to the number of Fock states given in Eq.\ (\ref{eq:choose}).
The $N$ replicas of condition (\ref{eq:norm}), in principle, reduce the number of independent real parameters by $N$ but we did not make use of this fact
in the numerics to be presented in the next section (although we monitored that (\ref{eq:norm}) is fulfilled).}
We stress that the Ansatz above goes beyond mean-field Ans\"atze a la Gutzwiller, used, e.g., in \cite{Jaksch98}.

The time-dependent variational principle can be formulated in terms of the Lagrangian
\be
L= \frac{\rm i}{2}[\langle \Psi|\dot{\Psi}\rangle-\langle\dot{\Psi}|\Psi\rangle]-\langle \Psi(t)|\hat{H}|\Psi(t)\rangle,
\ee
and is leading to the Euler-Lagrange equations
\be
\label{eq:EL}
\frac{\partial L}{\partial u_k^*}-\frac{\rm d}{{\rm d}t}\frac{\partial L}{\partial \dot{u}_k^*}=0,
\ee
where $u_k$ can be any element of  $\{A_k,\xi_{k1},\cdots\xi_{kM} \}$ \cite{KS81}. A
comparison of the different variational principles in this context, which are McLachlan, time-dependent and Dirac-Frenkel is performed in  \cite{BLKvL88} and \cite{LF14,ASC08} as well as in \cite{Werther2020}.
From Eq. (\ref{eq:var}) it is obvious that the wavefunction Ansatz does not contain the
complex conjugates of any of the parameters and that thus the Cauchy-Riemann conditions
are fulfilled and all three variational principles are equivalent, \textcolor{black}{analogous to the equivalence worked out in \cite{ASC08} for MCTDHB.}

To proceed, we first need to calculate the expression (suppressing the $S$ dependence of the basis states in
a short-hand notation)
\begin{align}\label{multiT}
 \frac{\rm i}{2}[\langle \Psi|\dot{\Psi}\rangle-\langle\dot{\Psi}|\Psi\rangle]\non &=\frac{\rm i}{2}\sum_{k,j=1}^N(A_k^*\dot{A}_j-\dot{A}^*_kA_j)\langle \vec{\xi}_k|\vec{\xi}_j\rangle\\
 &+\frac{\rm i}{2}S\sum_{k,j=1}^NA_k^*A_j\sum_{i=1}^M(\xi_{ki}^*\dot{\xi}_{ji}-\dot{\xi}_{ki}^*\xi_{ji})\langle\vec{\xi^{'}_k}|\vec{\xi^{'}_j}\rangle
\end{align}
for the time-derivative. Here we have used the expressions for the right and left time derivatives from Eqs.\ (\ref{righttime},\ref{lefttime}) to
arrive at  the final result.

Furthermore, from the Hamiltonian for the multimode BH model in Eq. (\ref{eq:BH}), we get the expectation value
\begin{align}\label{multiH}
  H \non &=\langle \Psi(t)|\hat{H}|\Psi(t)\rangle\\ \non &=\sum_{k,j=1}^NA_k^*A_j\Big[-JS\sum_{i=1}^{M-1}(\xi_{ki}^*\xi_{j,i+1}+\xi_{k,i+1}^*\xi_{ji})\langle\vec{\xi^{'}_k}|\vec{\xi_j^{'}}\rangle+\frac{U}{2}S(S-1)\sum_{i=1}^M\xi_{ki}^{*2}\xi_{ji}^2\langle\vec{\xi_k^{''}}|\vec{\xi_j^{''}}\rangle\\
  &+S\frac{K}{2}\sum_{i=1}^M(i-j_0)^2\xi_{ki}^*\xi_{ji}\langle\vec{\xi^{'}_k}|\vec{\xi^{'}_j}\rangle\Big]
\end{align}
where we have again used definitions of the $(S-1)$ and $(S-2)$-boson GCS from Appendix \ref{app:prop}.
Combining Eq.~(\ref{multiT}) and Eq.~(\ref{multiH}), we arrive at the Lagrangian for the multimode BH model
\begin{align}
 L\non &=\frac{\rm i}{2}\sum_{k,j=1}^N(A_k^*\dot{A}_j-\dot{A}^*_kA_j)\langle \vec{\xi}_k|\vec{\xi}_j\rangle+\frac{\rm i}{2}S\sum_{k,j=1}^NA_k^*A_j\sum_{i=1}^M(\xi_{ki}^*\dot{\xi}_{ji}-\dot{\xi}_{ki}^*\xi_{ji})\langle\vec{\xi^{'}_k}|\vec{\xi^{'}_j}\rangle\\
 \non &-\sum_{k,j=1}^NA_k^*A_j\Big[-JS\sum_{i=1}^{M-1}(\xi_{ki}^*\xi_{j,i+1}+\xi_{k,i+1}^*\xi_{ji})\langle\vec{\xi^{'}_k}|\vec{\xi_j^{'}}\rangle+\frac{U}{2}S(S-1)\sum_{i=1}^M\xi_{ki}^{*2}\xi_{ji}^2\langle\vec{\xi_k^{''}}|\vec{\xi_j^{''}}\rangle\\
 &+S\frac{K}{2}\sum_{i=1}^M(i-j_0)^2\xi_{ki}^*\xi_{ji}\langle\vec{\xi^{'}_k}|\vec{\xi^{'}_j}\rangle\Big].
\end{align}
This leads to the following derivatives of the Lagrangian with respect to the coefficients $A_k^*$ and their time derivatives
\begin{align}
 \frac{\partial L}{\partial A^*_k}\non \non &=\frac{\rm i}{2}\sum_{j=1}^N\dot{A}_j\langle \vec{\xi_k}|\vec{\xi_{j}}\rangle+\frac{\rm i}{2}S\sum_{j=1}^NA_j\sum_{i=1}^M(\xi_{ki}^*\dot{\xi}_{ji}-\dot{\xi}_{ki}^*\xi_{ji})\langle\vec{\xi^{'}_k}|\vec{\xi^{'}_j}\rangle-\frac{\partial H}{\partial A_k^*}\\
 \frac{\rm d}{{\rm d}t}\frac{\partial L}{\partial \dot{A}_k^*}&=-\frac{\rm i}{2}\sum_{j=1}^N\dot{A}_j\langle \vec{\xi}_k|\vec{\xi}_j\rangle-\frac{\rm i}{2}S\sum_{j=1}^NA_j\sum_{i=1}^M(\xi^{*}_{ki}\dot{\xi}_{ji}+\dot{\xi}_{ki}^*\xi_{ji})\langle\vec{\xi_k^{'}}|\vec{\xi^{'}_j}\rangle,
\end{align}
where
\begin{align}
  \frac{\partial H}{\partial A_k^*} \non &=\sum_{k,j=1}^NA_j\Big[-JS\sum_{i=1}^{M-1}(\xi_{ki}^*\xi_{j,i+1}+\xi_{k,i+1}^*\xi_{ji})\langle\vec{\xi^{'}_k}|\vec{\xi_j^{'}}\rangle+\frac{U}{2}S(S-1)\sum_{i=1}^M\xi_{ki}^{*2}\xi_{ji}^2\langle\vec{\xi_k^{''}}|\vec{\xi_j^{''}}\rangle\\
  &+S\frac{K}{2}\sum_{i=1}^M(i-j_0)^2\xi_{ki}^*\xi_{ji}\langle\vec{\xi^{'}_k}|\vec{\xi^{'}_j}\rangle\Big]
\end{align}
From the Euler Lagrange equation (\ref{eq:EL}), the equation of motion for the coefficient $A_j$
\begin{align}\label{Ak}
{\rm i}\sum_{j=1}^N\dot{A}_j\langle \vec{\xi_k}|\vec{\xi_{j}}\rangle+iS\sum_{j=1}^NA_j\sum_{i=1}^M\xi_{ki}^*\dot{\xi}_{ji}\langle\vec{\xi^{'}_k}|\vec{\xi^{'}_j}\rangle-\frac{\partial H}{\partial A_k^*}=0
\end{align}
can thus be obtained.

Next we switch to the equations of motion for the parameters $\xi_{km}$ of the SU($M$) coherent states. For the corresponding derivatives of the Lagrangian, we find
\begin{align}
 \frac{\partial L}{\partial \xi^*_{km}}\non &=\frac{\rm i}{2}S\sum_{j=1}^N(A_k^*\dot{A}_j-\dot{A}^*_kA_j)\xi_{jm}\langle\vec{\xi_k^{'}}|\vec{\xi_j^{'}}\rangle+\frac{\rm i}{2}S\sum_{j=1}^NA_k^*A_j\dot{\xi}_{jm}\langle\vec{\xi_k^{'}}|\vec{\xi_j^{'}}\rangle\\
 &+\frac{\rm i}{2}S(S-1)\sum_{j=1}^NA_k^*A_j\Big[\sum_{i=1}^M(\xi_{ki}^*\dot{\xi}_{ji}-\dot{\xi}_{ki}^*\xi_{ji})\xi_{jm}\langle\vec{\xi_k^{''}}|\vec{\xi_j^{''}}\rangle\Big]-\frac{\partial H}{\partial \xi_{km}^*}\\
 \frac{\rm d}{{\rm d}t}\frac{\partial L}{\partial \dot{\xi}_{km}^*}\non &=-\frac{\rm i}{2}S\sum_{j=1}^N(\dot{A}_k^*A_j+A_k^*\dot{A}_j)\xi_{jm}\langle\vec{\xi_k^{'}}|\vec{\xi_j^{'}}\rangle-\frac{\rm i}{2}S\sum_{j=1}^NA_k^*A_j\dot{\xi}_{jm}\langle\vec{\xi_k^{'}}|\vec{\xi_j^{'}}\rangle\\
 &-\frac{\rm i}{2}S(S-1)\sum_{j=1}^NA_k^*A_j\sum_{j=1}^M(\dot{\xi}_{ki}^*\xi_{ji}+\xi_{ki}^*\dot{\xi}_{ji})\xi_{jm}\langle\vec{\xi_k^{''}}|\vec{\xi_j^{''}}\rangle
\end{align}
with
\begin{align}
\frac{\partial H}{\partial \xi_{km}^*}\non &=\sum_{j=1}^NA_k^*A_j\Big[-JS(\xi_{j,m+1}+\xi_{j,m-1})\langle\vec{\xi_k^{'}}|\vec{\xi_j^{'}}\rangle\\
\non &-JS(S-1)\sum_{i=1}^{M-1}(\xi_{ki}^*\xi_{j,i+1}+\xi_{k,i+1}^*\xi_{j,i})\xi_{jm}\langle\vec{\xi_k^{''}}|\vec{\xi_j^{''}}\rangle\\
\non &+US(S-1)\xi_{km}^*\xi_{jm}^2\langle\vec{\xi_k^{''}}|\vec{\xi_j^{''}}\rangle+\frac{U}{2}S(S-1)(S-2)\sum_{i=1}^M\xi_{ki}^{*2}\xi_{ji}^2\xi_{jm}\langle\vec{\xi_k^{'''}}|\vec{\xi_j^{'''}}\rangle\\
&+\frac{K}{2}S(m-j_0)^2\xi_{jm}\langle\vec{\xi_k^{'}}|\vec{\xi_j^{'}}\rangle+\frac{K}{2}S(S-1)\sum_{i=1}^M(i-j_0)^2\xi_{ki}^*\xi_{ji}\xi_{jm}\langle\vec{\xi_k^{''}}|\vec{\xi_j^{''}}\rangle\Big]
\end{align}
Again using the Euler-Lagrange equation, we finally arrive at the differential equation
\begin{align}\label{xikm}
 &{\rm i}S\Bigl[\sum_{j=1}^NA_k^*\dot{A}_j\xi_{jm}\langle\vec{\xi_k^{'}}|\vec{\xi_j^{'}}\rangle+\sum_{j=1}^NA_k^*A_j\dot{\xi}_{jm}\langle\vec{\xi_k^{'}}|\vec{\xi_j^{'}}\rangle+(S-1)\sum_{j=1}^NA_k^*A_j\sum_{i=1}^M\xi_{ki}^*\dot{\xi}_{ji}\xi_{jm}\langle\vec{\xi_k^{''}}|\vec{\xi_j^{''}}\rangle\Bigr]\\
 \non&-\frac{\partial H}{\partial \xi_{km}^*}=0
\end{align}
for the $\xi_{km}$. Eqs.\ (\ref{Ak},\ref{xikm}) are coupled and nonlinear,
as in the more standard case of Glauber coherent states as basis functions. There are no exponential function nonlinearities though, because the GCS overlaps are ``simpler'' than in the Glauber case,
as can be seen in Appendix \ref{app:prop}.

In Appendix \ref{app:matrix}, we put the nonlinear, coupled, implicit differential equations for the GCS parameters and
the expansion coefficients in matrix form, which
allows us to solve them efficiently numerically.

\section{Numerical results}

As a proof of principle, in the following, we will present numerical results for the population dynamics of the BH model with $J=1$, based on the complex grid for SU($M$) coherent states introduced in Sect.\ \ref{ssec:disc}
for increasing mode number $M< S$, leading to an ever increasing size of the Fock state basis. \textcolor{black}{If possible, numerical results from a calculation in the full Fock space are compared with
the GCS results. As a guide, in Table \ref{tab:para}, the different parameters for which results
will be presented are collected, highlighting the dimension of the Fock space as well as the number of complex parameters, $(M+1)N$, needed for converged results using the GCS method.
We note that almost converged results can be gained by using much smaller GCS bases than indicated in the table. Furthermore, for longer time series more basis functions will be needed and,
as shown in Appendix \ref{app:grid}, closely spaced grids help to keep the number of basis functions needed for convergence small. For minimization of the number of basis functions, both, randomization
of the grid, as well as reduction of grid spacing should be performed. This has been done only for the cases $M=4$ and $M=6$, however.}
\begin{table}[h!]
  \begin{center}
    \label{tab:para}
    \begin{tabular}{r||c|c|c|c} 
      & 2   & 3    & 4   & 6\\
      \hline
      \hline
      20 &     & $C^F$=231 &     & $C^F$=53130\\
      &&200, \ref{ssec:mode3}&&3500, \ref{ssec:mode6} \\
      \hline
      30 &     &      &$C^F$=5456  &  \\
      &&&500, App.~\ref{app:grid},&\\
      \hline
      50 &$C^F$=51& & & \\
      &75, \ref{ssec:mode2},  App.~\ref{app:grid} &&&\\
      \hline
      200 &$C^F$=201& & & \\
      &243, App.~\ref{app:grid} &&&\\
    \end{tabular}
        \caption{Collection of parameters of numerical studies presented below. In dependence of mode number $M$ (column index) and particle number $S$ (line index), the table's entries give the size of the Fock state basis from Eq.\ (\ref{eq:choose}),
        as well as the number of complex parameters needed for converged GCS results and where in the text the numerical results can be found. For  $M=2,S=200$, the underlying grid was a diagonal grid (see App.\ \ref{app:grid}).
        For $M=2,S=50$, the grid in Sec.\ \ref{ssec:mode2} was random with spacing $\sqrt{\pi}$ and the one in App.\ \ref{app:grid} was diagonal but with optimized spacing.}
  \end{center}

\end{table}

\subsection{The two mode case $M=2$}
\label{ssec:mode2}

Firstly, for the driven two-mode BH model with a time-dependent tunneling matrix element studied in \cite{TWK09}, the Hamiltonian can be written as
\begin{align}
H=-J(t)(a_1^\dag a_2+a_2^\dag a_1)+\frac{U}{2}(a_1^{\dag 2}a_1^{2}+a_2^{\dag 2}a_2^{2})+\frac{K}{2}\sum_{j=1}(j-j_0)^2a_j^\dag a_j
\end{align}
where $J(t)=J_0+J_1\cos(\omega t)$. The initial state
\be
|\Psi(0)\rangle=|S=50, \xi_1=-\sqrt{0.7},\xi_2=\sqrt{0.3}\rangle
\ee
used for the propagation is the same as in \cite{TWK09}.

By means of the SU($2$) coherent states originating from the complex grid of Glauber coherent with distance $\sqrt{\pi}$, we employ the Ansatz
\be
|\Psi(t)\rangle=\sum_{k=1}^NA_k(t)|S=50,\xi_{k1}(t),\xi_{k2}(t)\rangle
\ee
for the wavefunction that contains 3$N$ complex parameters. \textcolor{black}{The random grid one is constructed such that it
contains the initial state in the ``middle'', as indicated in Fig.\ \ref{compg}.}

\begin{figure}[h]
  \includegraphics[width=3.4in]{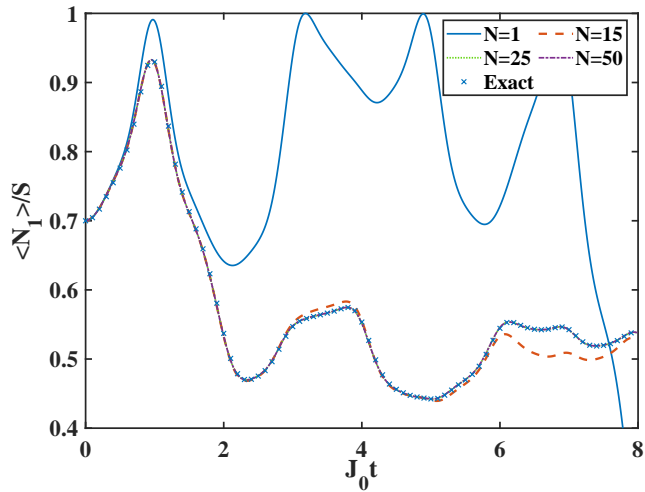}
  \caption{Dynamics of the population of the first mode of a two mode BH model with different numbers of SU(2) basis states.
  The particle number is $S=50$, the hopping parameter is $J_0=1$, the on-site interaction energy is $U=0.1J_0$ and $K=0$.
  The driving frequency and strength are $\omega=2\pi/J_0$ and $J_1=0.5J_0$.
  Different lines are results for different numbers of GCS: solid: $N=1$, dashed: $N=15$, dotted: $N=25$, dash-dotted: $N=50$.
  \textcolor{black}{Stars denote Fock space results.}}
  \label{fig:mode2}
\end{figure}

Converged results for the  population \textcolor{black}{(normalized to the total number of particles) of the first mode as a function of time} were obtained by using $N=$25 basis states and even 15 states lead to almost perfect results as is shown in Fig.~(\ref{fig:mode2}).
These results agree with the ones shown in Fig.\ 12(c) in \cite{TWK09}. We stress that for trivial multiplicity $N=1$, the result is only
reasonable for a very short time period and thereafter looses its predictive power, as can be seen in Fig.~(\ref{fig:mode2}).

There is no big difference in the number of complex parameters that have been used for the converged results, compared to the size of the Fock space basis,
which here is just 51, however. Therefore let us investigate increasing mode number in the following.

\subsection{The three mode case $M=3$}
\label{ssec:mode3}

\begin{figure}
  \centering
  \includegraphics[width=3.4in]{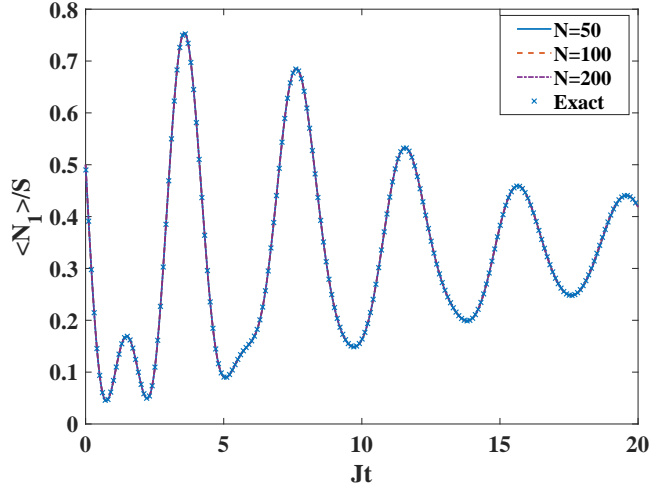}\\
  \caption{Dynamics of the population of the first mode for three mode case for initial
  state (\ref{eq:ini}). The on-site interaction energy is $U=0.1J$, whereas $K=0$. Results
  for different number of basis functions (solid: $N=50$, dashed $N=100$, dash-dotted $N=200$)
  are almost  indistinguishable. \textcolor{black}{Stars denote Fock space results.} }\label{mode3}
\end{figure}

For three modes occupied by 20 bosons, we first choose the superposition of Fock states
\be
|\Psi(0)\rangle=|S=20,\xi_1=\frac{\sqrt{2}}{2},\xi_2=i\frac{\sqrt{2}}{2},\xi_3=0\rangle
\label{eq:ini}
\ee
as the initial state.
In this case, our numerical results in Fig.~(\ref{mode3}) show that convergence is achieved
with 50 basis functions, i.e. 200 complex parameters, a little less than the 231 complex-valued
amplitudes one would need for an expansion in Fock states in the present case.

\begin{figure}[htbp]
  \includegraphics[width=3.4in]{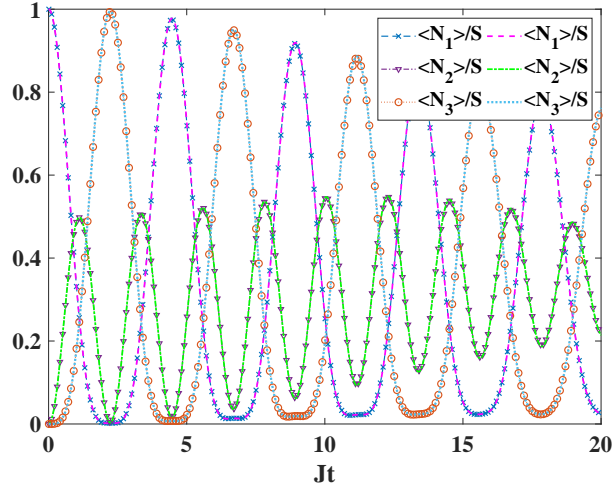}\\
  \caption{Dynamics of the population of all three modes. The initial state is the Fock state
  $|20,1,0,0\rangle$. $U=0.03J$ and $\Lambda=0.6$ is in the Rabi regime. The exact results are displayed by the different markers (crosses: $\langle N_1\rangle/S$, triangles: $\langle N_2\rangle /S$, circles: $\langle N_3\rangle /S$),
  while the corresponding lines without markers are calculated by variational dynamics using 50 basis functions whose distance is $\sqrt\pi$.}
  \label{fig:Rab}
\end{figure}

\begin{figure}[htbp]
  \includegraphics[width=3.4in]{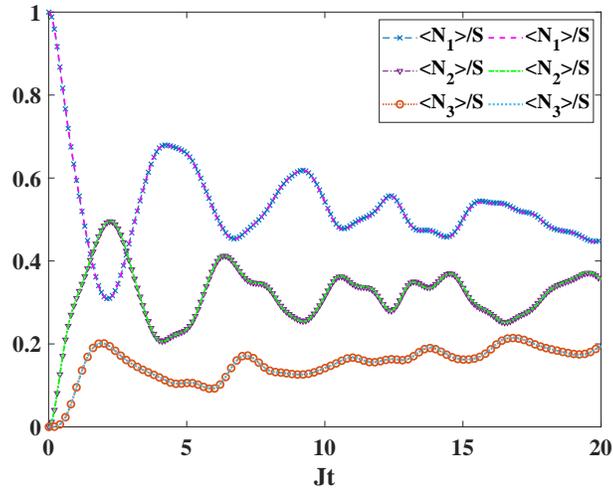}\\
  \caption{Dynamics of the population of all three modes. The initial state is the Fock state
  $|20,1,0,0\rangle$. $U=0.2J$ and $\Lambda=4$ is in the Josephson regime. The exact results are displayed by the different markers (crosses: $\langle N_1\rangle/S$, triangles: $\langle N_2\rangle /S$, circles: $\langle N_3\rangle /S$), while the corresponding lines without markers
  are calculated by the variational dynamics using 50 basis functions whose distance is $\sqrt\pi$.}
  \label{fig:mid}
\end{figure}

The parameter combination $\Lambda={US/J}$ is frequently applied to distinguish the
dynamical features of the BH model. For $\Lambda<1$  the dynamics is located in Rabi regime, while $1<\Lambda\ll S^2$ and $\Lambda\gg S^2$ represent the so-called Josephson and Fock regime, respectively \cite{Leg01,SiSt14}.
In Figs.\ \ref{fig:Rab} and \ref{fig:mid}, we show the  accuracy of the variational dynamics for the initial Fock state
\be
|\Psi(0)\rangle=|S=20,1,0,0\rangle
\ee
by comparing with the exact numerical results
(gained by an expansion of the wavefuncion in terms of Fock states) for 3 modes and 20 bosons. It turns out that the 50 basis functions that were found sufficient
in Fig.\ \ref{mode3} are still sufficient in the Rabi and Josephson cases. The pure Fock regime case was too demanding numerically and is therefore not considered.
In the Rabi case displayed in Fig.\ \ref{fig:Rab} almost undamped sinusoidal oscillations are observed.
The oscillations become more complex and damped in the Josephson case shown in Fig.\ \ref{fig:mid}. In this case, also the accuracy of ode45 \cite{MATLAB:2020} used for the solution of
the differential equations had to be promoted beyond the default value.

\subsection{Beyond $M=3$}
\label{ssec:mode6}

To highlight the power of the proposed method we now investigate the more complicated situation
where $20$ bosons occupy $6$ modes. Then the Hilbert space consists of ${25 \choose 5}=53130$
Fock states \textcolor{black}{and with a standard desktop computer, it will be difficult to propagate in this large Hilbert space using the full Fock state basis.}

\begin{figure}[htbp]
  \centering
  \includegraphics[width=4in]{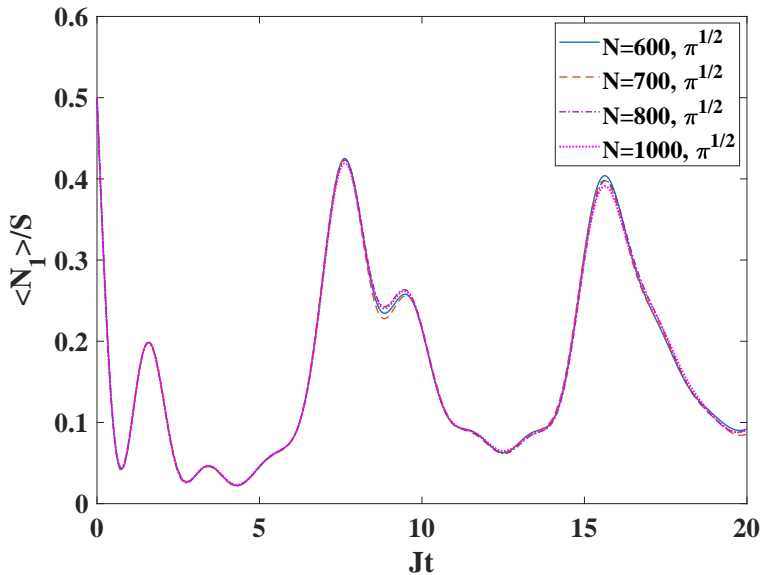}\\
  \caption{The dynamics of population of the first mode for the case of six modes. The initial state is $|S=20,\xi_1=\frac{\sqrt{2}}{2},\xi_2=i\frac{\sqrt{2}}{2},\xi_3=\cdots\xi_6=0\rangle$.
  The on-site interaction energy is $U=0.1J$ and $K=0$. Solid line: 600 basis functions, dashed line: 700 basis functions, dashed-dotted line: 800 basis functions, dotted line: 1000 basis functions. The
  spacing of the underlying Glauber CS grid was $\sqrt{\pi}$.}\label{mode6_1}
\end{figure}

If we employ the SU$(M)$ states originating from the complex grid with the von Neumann spacing $\sqrt{\pi}$, as presented in Fig.~(\ref{mode6_1}), we find that
only several hundreds of GCS are needed for converged results at longer times. This means we need $(M+1)\times 800=5600$ complex valued parameters in our Ansatz (\ref{eq:var}),
in order to converge the results for all times shown. Especially at the extrema of the curve, a large number of basis functions is needed.

\begin{figure}[htbp]
  \centering
  \includegraphics[width=3.4in]{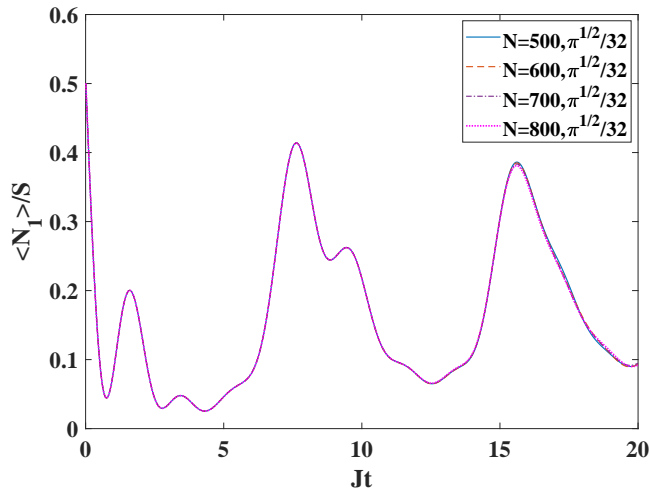}\\
  \caption{Convergence of the results for a distance of $\frac{\sqrt{\pi}}{32}$ of the underlying Glauber CS grid.
  Other parameters (apart from the number of basis functions) as in Fig.~(\ref{mode6_1})}\label{mode6_3}
\end{figure}

However, if we decrease the distance of the complex grid, the results can be converged faster. In Fig.~(\ref{mode6_3}), we show that the underlying grid with the smaller distance of
$\sqrt{\pi}/32$ between the grid points
allows for convergence already with 500 basis functions, whereas the ``standard'' grid needed 800 states to converge the results at specific later times.
\textcolor{black}{An in-depth study of the usage of
denser underlying grids is given in Appendix \ref{app:grid}.}
\\\\\\
These numerical results show that a GCS with only $M$ complex parameters according to Eq. (\ref{wh}) and a corresponding variational Ansatz (\ref{eq:var}) with $(M+1)N$ parameters
allows for a faithful description of the dynamics under the BH Hamiltonian. The number of independent complex parameters that we had to employ is more than one order of magnitude smaller than the dimension
of the complete Fock space.

\section{Conclusions and Outlook}

We have shown that a numerically favorable treatment of BH dynamics in 1D
is possible by using a variational approach based on GCS.
In contrast to established mean-field versions of the theory, here
we have employed a numerically complete set of SU($M$) GCS
as basis states in the expansion of the initial wavefunction. We have built an overcomplete
set of basis function by using the expansion of the SU($M$) states in terms of Glauber
coherent states for whom it is well-established how a numerically complete set of basis states
has to be chosen.

The equations of motion of the GCS parameters as well as of the expansion coefficients have been
derived from the time-dependent variational principle in its Lagrangian form.
The central idea is that the number of GCS parameters
times the number of basis functions is much smaller than the number of basis functions needed in an
unbiased expansion of the wavefunction in terms of Fock states. In this respect the method
is closely related to the multi-configuration time-dependent Hartree method for bosons \cite{ASC08}
as well as the variational approach in \cite{RMW10,RMCKW10}, which
are both, however, not based on GCS.

That the method indeed works for considerably smaller numbers of basis functions than in the
case of an expansion in terms of Fock states has shown to be true for large particle number $S$ and increasing mode number $M<S$ for several examples in different parameter ranges.
Especially for larger mode number, the use of small spacing (considerably smaller than the von Neumann spacing of $\sqrt\pi$)  in the underlying Glauber CS
grid turned out to be necessary for the use of a relatively small number of GCS basis functions. In this way we could report converged results using more than one order of magnitude less
parameters than would have been necessary for an expansion in terms of Fock states.
We stress that we did not encounter any problems that had to be solved
with the recently introduced apoptosis procedure \cite{prb20}. Regularization also used in
\cite{prb20} and well known from other approaches like MCTDH \cite{Manthe1992} was necessary, however.

In the future, we plan to investigate also larger mode numbers and small particle
numbers that are relevant in thermalization studies of cold atoms in optical
superlattices \cite{Tetal12}, as well as other particle number conserving bosonic lattice systems with the proposed method.
If its fortunate scaling properties persist, new territory in parameter space may become explorable \textcolor{black}{and we are especially interested in the case where $U$ is much larger than $J$}.
Furthermore, the application of the proposed methodology to BH models in more than 1D is planned. It is well known that the entanglement entropy
of the ground state in lattice models fulfills so-called area laws \cite{ECP10}.
The entanglement growth in the course of time evolution poses a serious problem for other numerical propagation techniques like
matrix product state based methods, however \cite{Scholl2011}.
It remains to be explored how well the method proposed herein can cope with these problems
in higher dimensions.

{\bf Acknowledgments:}
The authors would like to thank Valentin Link for enlightening discussions.
FG would like to thank the Deutsche Forschungsgemeinschaft for financial support under grant GR 1210/8-1.

\begin{appendix}

\section{Some properties of SU($M$) coherent states}
\label{app:prop}

In this appendix, we review some computationally helpful formulae along the
lines of \cite{BP08}, which are needed in the main text. Firstly, the commutation between the annihilation operator and the collective creation operator which is applied to
generate an SU($M$) coherent state is given by
\begin{align}\label{commutation}
[\hat a_i,\Big(\sum_{j=1}^M\xi_j\hat a_j^\dag\Big)^S]=S\xi_i\Big(\sum_{j=1}^M\xi_j\hat a_j^\dag\Big)^{S-1}.
\end{align}
Secondly, by defining two collective operators
\begin{align}
 \hat A^\dag=\sum_{i=1}^M\xi_i\hat a_i^\dag,\indent \hat B^\dag=\sum_{i=1}^M\eta_i\hat a_i^\dag,
\end{align}
two different SU($M$) coherent states can be generated via
\begin{align}
 |\vec{\xi}\,\rangle=\frac{1}{\sqrt{S!}}(\hat A^\dag)^S|0\rangle,\indent
 |\vec{\eta}\,\rangle=\frac{1}{\sqrt{S!}}(\hat B^\dag)^S|0\rangle.
\end{align}
The vectorized parameter contains the parameters of all modes (here we do not
consider doubly indexed parameters).
The inner product of the above two states is then given by
\begin{align}
 \langle \vec{\eta}|\vec{\xi}\,\rangle\non &=\frac{1}{S!}\langle 0|\Big(\sum_{i=1}^M\eta^*_i \hat a_i\Big)^S\Big(\sum_{j=1}^M\xi_j\hat a_j^\dag\Big)^S|0\rangle\\
 \non &=\frac{1}{S!}\langle 0|\sum_{n_1+n_2+\cdots=S}\frac{S!}{n_1!n_2!\cdots}[(\eta_1^*\hat a_1)^{n_1}(\eta_2^*\hat a_2)^{n_2}\cdots]\\
 \non&\sum_{m_1+m_2+\cdots=S}\frac{S!}{m_1!m_2!\cdots}[(\xi_1\hat a_1^\dag)^{m_1}(\xi_2\hat a_2^\dag)^{m_2}\cdots]|0\rangle
 \non\\&=\frac{1}{S!}\Big(\sum_{n_1+n_2+\cdots=S}\frac{S!}{\sqrt{n_1!n_2!\cdots}}\eta_1^{*n_1}\eta_2^{*n_2}\cdots\langle \vec{n}|\Big)
 \non\\&\sum_{m_1+m_2+\cdots=S}\frac{S!}{\sqrt{m_1!m_2!\cdots}}\xi_1^{m_1}\xi_2^{m_2}\cdots|\vec{m}\rangle
 \non\\
 \non &=\sum_{m_1+m_2+\cdots=S}\frac{S!}{m_1!m_2!\cdots}(\eta^*_1\xi_1)^{m_1}(\eta_2^*\xi_2)^{m_2}\cdots\\
  &=\Big(\sum_{i=1}^M\eta_i^*\xi_i\Big)^S,
  \label{eq:ip}
\end{align}
where we have used the general binomial theorem
\begin{align}
 (x_1+x_2+\cdots+x_n)^k=\sum_{a_1+a_2+\cdots+a_n=k}\frac{k!}{a_1!a_2!\cdots a_n!}x_1^{a_1}x_2^{a_2}\cdots x_n^{a_n}.
\end{align}
The above result (\ref{eq:ip}) is quite different from the corresponding property for Glauber coherent states, which involves exponential functions.

Using Eq.~(\ref{commutation}), we can now calculate the action of the annihilation operator on
the SU($M$) coherent state via
\begin{align}\label{annihilation}
 \hat a_i|\vec{\xi}\,\rangle\non &=\hat a_i\frac{1}{\sqrt{S!}}\Big(\sum_{j=1}^M\xi_j\hat a_j^\dag\Big)^S|0\rangle\\
 \non &=\frac{1}{\sqrt{S!}}[\Big(\sum_{j=1}^M\xi_j\hat a_j^\dag\Big)^Sa_i+S\xi_i\Big(\sum_{j=1}^M\xi_j\hat a_j^\dag\Big)^{S-1}]|0\rangle\\
 &=\sqrt{S}\xi_i|\vec{\xi^{'}}\rangle
\end{align}
where we have used the action of the annihilation operator on the ground state, see
Eq.\ (\ref{eq:creann}) as well as the definition
\be
|\vec{\xi^{'}}\rangle=\frac{1}{\sqrt{(S-1)!}}\Big(\sum_{j=1}^M\xi_j\hat a_j^\dag\Big)^{S-1}|0\rangle
\ee
of the $(S-1)$-boson GCS.

Furthermore, for $|\eta\rangle$ and $|\xi\rangle$, from Eq.\ (\ref{annihilation}) we get
\begin{align}\label{transition}
 \langle \vec{\eta}|\hat a_j^\dag \hat a_k|\vec{\xi}\,\rangle=S\eta^*_j\xi_k\langle \vec{\eta^{'}}|\vec{\xi^{'}}\rangle
\end{align}
where the inner product $\langle \vec{\eta^{'}}|\vec{\xi^{'}}\rangle$ is
\begin{align}
\label{eq:singleprime}
 \langle \vec{\eta^{'}}|\vec{\xi^{'}}\rangle=\Big(\sum_{i=1}^M \eta_i^*\xi_i\Big)^{S-1}.
\end{align}
In the main text as well as below, we will also refer to the following two inner products
\begin{align}
\label{eq:doubleprime}
 \langle \vec{\eta^{''}}|\vec{\xi^{''}}\rangle&=\Big(\sum_{i=1}^M \eta_i^*\xi_i\Big)^{S-2},
\\
 \langle \vec{\eta^{'''}}|\vec{\xi^{'''}}\rangle&=\Big(\sum_{i=1}^M \eta_i^*\xi_i\Big)^{S-3},
\end{align}
of the $(S-2)$ and the $(S-3)$ GCS.

Using the chain rule, the matrix element of the right time-derivative now is
\begin{align}\label{righttime}
 \langle \vec{\eta}|\overrightarrow{\partial_t}|\vec{\xi}\,\rangle\non &=\langle \vec{\eta}|\sum_{i=1}^M\dot{\xi_i}\partial_{\xi_i}|\vec{\xi}\,\rangle
 \\
 \non &=\frac{S}{\sqrt{S!}}\langle\vec{\eta}|\sum_{i=1}^M(\dot{\xi}_i\hat a_i^\dag)(\hat A^\dag)^{S-1}|0\rangle
 \\
 \non &=\frac{S}{S!}\langle0|\hat B^{S-1}\hat B\sum_{i=1}^M(\dot{\xi}_i\hat a_i^\dag)(\hat A^\dag)^{S-1}|0\rangle
 \\
 \non &=\langle \vec{\eta^{'}}|\sum_{i=1}^M(\dot{\xi}_i\hat B\hat a_i^\dag)|\vec{\xi^{'}}\rangle
 \\
 \non &=\langle \vec{\eta^{'}}|\sum_{i=1}^M\dot{\xi}_i(\eta_i^*+\hat a_i^\dag \hat B)|\vec{\xi^{'}}\rangle
 \\
 \non &=\sum_{i=1}^M(\dot{\xi}_i\eta^*_i)\langle\vec{\eta^{'}}|\vec{\xi^{'}}\rangle+\sum_{i,j=1}^M\dot{\xi}_i\eta^*_j\langle\vec{\eta^{'}}| \hat a_i^\dag \hat a_j|\vec{\xi^{'}}\rangle
 \\
 \non &=\sum_{i=1}^M(\dot{\xi}_i\eta^*_i)\langle\vec{\eta^{'}}|\vec{\xi^{'}}\rangle+(S-1)\sum_{i,j=1}^M\dot{\xi}_i\eta^*_j\eta_i^*\xi_j\langle\vec{\eta^{''}}|\vec{\xi^{''}}\rangle
 \\
 &=S\sum_{i=1}^M(\dot{\xi}_i\eta^*_i)\langle\vec{\eta^{'}}|\vec{\xi^{'}}\rangle
\end{align}
In the fifth line of the above equation, we have used the result of Eq.~(\ref{commutation}) for $S=1$, and the relation $\sum_{j=1}^M\xi_{j}\eta_{j}^*\langle\vec{\eta^{''}}|\vec{\xi^{''}}\rangle=\langle\vec{\eta^{'}}|\vec{\xi^{'}}\rangle$
which follows from Eqs.\ (\ref{eq:singleprime},\ref{eq:doubleprime}) is also used to get the result of the last line.

Similarly, for the left time-derivative we find
\begin{align}\label{lefttime}
 \langle \vec{\eta}|\overleftarrow{\partial_t}|\vec{\xi}\,\rangle\non &=\langle \vec{\eta}|\sum_{i=1}^M\dot{\eta^*_i}\partial_{\eta^*_i}|\vec{\xi}\rangle\\
 \non &=\langle \vec{\eta^{'}}|\sum_{i=1}^M(\dot{\eta_i^*}\hat a_i\hat A^\dag)|\vec{\xi^{'}}\rangle\\
 \non &=\sum_{i=1}^M(\dot{\eta^*_i}\xi_i)\langle \vec{\eta^{'}}|\vec{\xi^{'}}\rangle+(S-1)\sum_{i,j=1}^M(\dot{\eta^{*}_i}\xi_j\eta^*_j\xi_i)\langle \vec{\eta^{''}}|\vec{\xi^{''}}\rangle\\
 &=S\sum_{i=1}^M(\dot{\eta^*_i}\xi_i)\langle \vec{\eta^{'}}|\vec{\xi^{'}}\rangle
\end{align}
Both results will be used in the main text.

\section{Matrix form of the variational equations}
\label{app:matrix}

Combining the equation Eq.~(\ref{Ak}) and Eq.~(\ref{xikm}), we get a compact and scalable matrix equation for
the vector containing all coefficients $\{A_k\}$ and GCS parameters $\{\xi_{km}\}$
\begin{equation}\label{matrixEq}
 \left(
 \begin{array}{cc}
  \bm{X} & \bm{Y}\\
  \bm{Y}^\dag & \bm{Z}
 \end{array}
 \right)
  \left(
 \begin{array}{cc}
  \bm{\dot{A}}\\
  \bm{\dot{\xi}}
 \end{array}
 \right)=
 -{\rm i}\left(
  \begin{array}{cc}
  \bm{R}_1\\
  \bm{R}_2
 \end{array}
 \right),
\end{equation}
analogous to the procedure lined out in the appendix of \cite{prb20} for Glauber coherent states.

The block matrices are given by
\begin{align}
 \bm{X}_{kj}&=\langle \vec{\xi_k}|\vec{\xi_j}\rangle,
\\
  \bm{Y}&=S(\bm{\xi}_1^*,\bm{\xi}_2^*,\cdots,\bm{\xi}_M^*)\otimes\bm{A}^T
 \circ(\bm{1}_{1\times M}\otimes \bm{X}^{'}),
\\
  \bm{Z}&=\bm{1}_{M\times M}\otimes \bm{\rho}\circ\bm{F},
\end{align}
where the vector $\vec{\xi_k}$ is now indexed by the basis function discretization index and the vector $\bm{\xi}_m$, to be defined below, is indexed by the mode index.
Furthermore,
\begin{align}
   \bm{F}_{ij}&=S(S-1)\bm{X^{''}}\circ({\bm\xi}_j^*\cdot{\bm\xi}_i^T) (i\neq j),
   \\
  \bm{F}_{ii}&=S\bm{X^{'}}+S(S-1)\bm{X^{''}}\circ({\bm\xi}_i^*\cdot{\bm\xi}_i^T),
\end{align}
and where $\bm{1}_{m\times n}$ is an $m\times n$ matrix which only consists of ones,
and $\bm{X}^{'}_{kj}=\langle \vec{\xi'_k}|\vec{\xi'_j}\rangle$ and $\bm{X}^{''}_{kj}=\langle \vec{\xi''_k}|\vec{\xi''_j}\rangle$ are overlaps of
$(S-1)$ and $(S-2)$-boson GCS from the previous appendix, respectively, whereas  $\bm{\rho}_{kj}=A_k^*A_j$.
Furthermore, $\otimes$ denotes the tensor-product,
whereas $\circ$ denoted the Hadamard-product (element-wise multiplication) and $\cdot$ denotes the standard scalar product.

Furthermore, the vectors are defined as
\begin{equation}
 \bm{\dot{A}}=
 \left(
 \begin{array}{ccc}
  \dot{A}_1\\
  \dot{A}_2\\
  \vdots\\
  \dot{A}_N
 \end{array}
 \right),
  \bm{\dot{\xi}}=
 \left(
 \begin{array}{ccc}
  \bm{\dot{\xi}_1}\\
  \bm{\dot{\xi}_2}\\
  \vdots\\
  \bm{\dot{\xi}_M}
 \end{array}
 \right),
   \bm{\xi_m}=
 \left(
 \begin{array}{cccc}
  \xi_{1m}\\
  \xi_{2m}\\
  \vdots\\
  \xi_{Nm}
 \end{array}
 \right),
 \bm{R_1}=
 \left(
  \begin{array}{ccc}
  \frac{\partial H}{\partial A_1^*}\\
  \frac{\partial H}{\partial A_2^*}\\
  \vdots\\
  \frac{\partial H}{\partial A_N^*}
 \end{array}
 \right),
    \bm{R_2}=
 \left(
  \begin{array}{cccc}
  \frac{\partial H}{\partial \xi_{11}^*}\\
  \frac{\partial H}{\partial \xi_{21}^*}\\
  \vdots\\
  \frac{\partial H}{\partial \xi_{1M}^*}\\
  \frac{\partial H}{\partial \xi_{2M}^*}\\
  \vdots
 \end{array}
 \right)
\end{equation}
with $H=\langle \Psi|H|\Psi\rangle$, the expectation of the
Hamiltonian given in Eq.\ (\ref{multiH}).

The vectors on the right hand side of equation Eq.~(\ref{matrixEq}) are
\begin{align}
 \bm{R}_1\non &=-JS\sum_{i=1}^{M-1}\big[\bm{X}^{'}\cdot(\bm{A}\circ\bm{\xi_{i+1}})\circ\bm{\xi}_i^*+\bm{X}^{'}\cdot(\bm{A}\circ\bm{\xi}_i)\circ\bm{\xi}_{i+1}^*\big]
 \non\\
 &+\frac{U}{2}S(S-1)\sum_{i=1}^M\bm{X}^{''}\cdot(\bm{A}\circ\bm{\xi}_i\circ\bm{\xi}_i)\circ\bm{\xi}_i^{*}\circ\bm{\xi}_i^*+\frac{K}{2}S\sum_{i=1}^M(i-j_0)^2\bm{X}^{'}\cdot(\bm{A}\circ\bm{\xi}_i)\circ\bm{\xi}_i^*
\end{align}
as well as
\begin{align}
 \frac{\partial H}{\partial \bm{\xi}_m^*}\non &=-JS\bm{\rho}\circ\bm{X}^{'}\cdot(\bm{\xi}_{m+1}+\bm{\xi}_{m-1})
 \non\\
 &-JS(S-1)\sum_{i=1}^{M-1}\big[\bm{\rho}\circ\bm{X}^{''}\cdot(\bm{\xi}_m\circ\bm{\xi}_i)\circ\bm{\xi}^*_{i+1}+\bm{\rho}\circ\bm{X}^{''}\cdot(\bm{\xi}_m\circ\bm{\xi}_{i+1})\circ\bm{\xi}_i^*\big]
 \non\\
 &+US(S-1)\bm{\rho}\circ\bm{X}^{''}\cdot(\bm{\xi}_m\circ\bm{\xi}_m)\circ\bm{\xi}_m^{*}
 \non\\
 &+\frac{U}{2}S(S-1)(S-2)\sum_{i=1}^M\big[\bm{\rho}\circ\bm{X}^{'''}\cdot(\bm{\xi}_i\circ\bm{\xi}_i\circ\bm{\xi}_m)\circ(\bm{\xi}_i^{*}\circ\bm{\xi}_i^{*})\big]
 \non\\
 &+\frac{K}{2}S(m-j_0)^2\bm{\rho}\circ\bm{X}^{'}\cdot\bm{\xi}_m+\frac{K}{2}S(S-1)\sum_{i=1}^M\left[(i-j_0)^2\bm{\rho}\circ\bm{X}^{''}\cdot(\bm{\xi}_m\circ\bm{\xi}_i)\circ\bm{\xi}_i^*\right]
\end{align}
for use in the vector ${\bm R}_2$.
Please keep in mind that $m+1\leq M, m-1\geq 1$.

\section{Detailed convergence study with respect to grid spacing}
\label{app:grid}

To prove the advantage of small spacings of complex grids for getting converged results, for most of this appendix we choose the diagonal of the combination of the rectangular grids rather than the random choice of points from
the rectangular grids which have been used in
the main text. In diagonal grids, the parameters of the SU($M$) coherent state lie in the exact same positions for every complex subgrid, for example,
\be
\{\xi_{mn,1},\xi_{m n,2}\}=\{\frac{z_{mn,1}}{\sqrt{|z_{mn,1}|^2+|z_{mn,2}|^2}},\frac{z_{mn,2}}{\sqrt{|z_{mn,1}|^2+|z_{mn,2}|^2}}\},
\ee
in the case of 2 modes, where $m$, $n$ are the same(!) complex grid indices.
We remind of the fact that in the random choice of the two-mode grid, a random combination of the points on the grid of the first mode with the points of the grid of the second mode is allowed!
Although using the diagonal grids is less optimal than the random ones (as will become obvious below), according to our numerical results, it can help us avoid the influence of randomness and compare the effects of different
spacings directly at the same level. In the following, we will show some results for normalized population dynamics for different initial states with different photon and mode numbers.
\begin{figure}[h]
  \centering
  \includegraphics[width=3in]{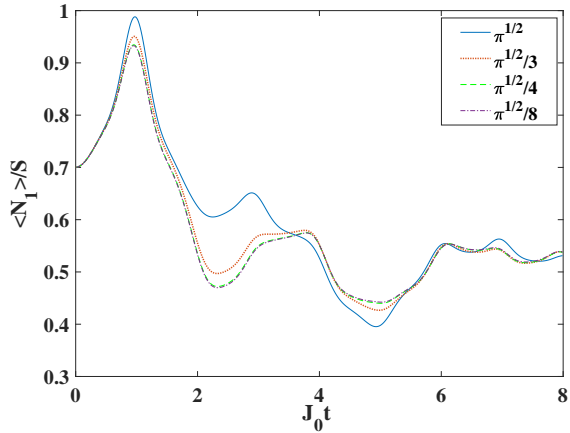}\\
  \caption{The dynamics of the population of the first mode in the two-mode case with 50 photons. The driving frequency and
  strength are $\omega=2\pi/J_0$ and $J_1=0.5J_0$, and the on-site interaction energy $U=0.1J$. The number of SU($M$) coherent states is 25.
  Different lines are results for different spacings of the complex grid. Solid: $\sqrt{\pi}$, dotted: $\sqrt{\pi}/3$, dash: $\sqrt{\pi}/4$, dash-dotted: $\sqrt{\pi}/8$.}\label{mode2_50_sg}
\end{figure}

\begin{figure}[h]
  \centering
  \includegraphics[width=3in]{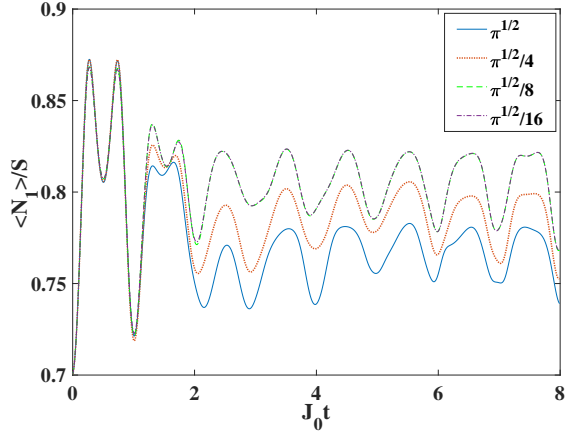}\\
  \caption{The dynamics of the population of the first mode in the two-mode case with 200 photons. The driving frequency and
  strength are $\omega=2\pi/J_0$ and $J_1=0.5J_0$, and the on-site interaction energy $U=0.1J$. The number of SU($M$) coherent states is 81. Different lines are results for different spacings of the complex grid:
  solid: $\sqrt{\pi}$, dotted: $\sqrt{\pi}/4$, dash: $\sqrt{\pi}/8$, dash-dotted: $\sqrt{\pi}/16$.}\label{mode2_200_sg}
\end{figure}

For the two-mode case, firstly, as in the main text, the initial state is chosen as $|50,-\sqrt{0.7},\sqrt{0.3}\rangle$ and the parameters of the Hamiltonian are also consistent with the ones used in Sect.\ \ref{ssec:mode2}.
We constructed two identical complex 5$\times$5 grids, and the number of SU($M$) we used thus is just $N=25$ (and not 625, as it would be if we would allow all combinations of the rectangular grid points). Fig.~(\ref{mode2_50_sg}) reveals that small spacings allow to arrive at the converged result more quickly.
In the present case, $\sqrt{\pi}/4$ is enough to reproduce the exact result. Please note that the result with 25 basis functions and the largest spacing of $\sqrt{\pi}$ is much worse than the one presented in Fig.\ \ref{fig:mode2},
where a random grid was used.
For the second, more demanding initial state $|200,-\sqrt{0.7},\sqrt{0.3}\rangle$ with 200 photons, the number of grid points is increased to 81. Fig.~(\ref{mode2_200_sg}) shows that the spacing with $\sqrt{\pi}/4$ does not yet
work well and that for this initial condition $\sqrt{\pi}/8$ is a better choice.

\begin{figure}[h]
  \centering
  \includegraphics[width=3in]{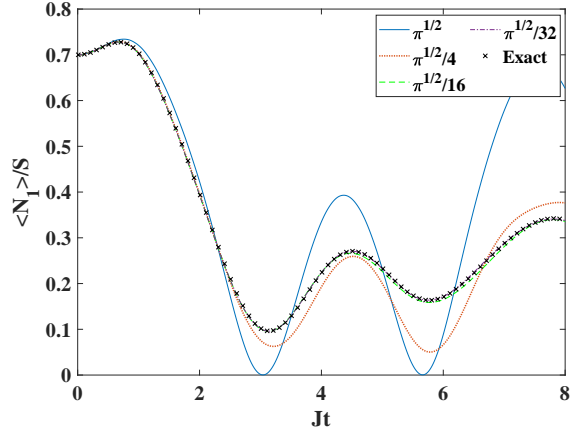}\\
  \caption{The dynamics of the population of the first mode in the four-mode case with 30 photons. The on-site interaction energy $U=0.1J$. The number of SU($M$) coherent states used is 169. Different lines are results for different spacings of the complex grid:
  solid: $\sqrt{\pi}$, dotted: $\sqrt{\pi}/4$, dash: $\sqrt{\pi}/16$, dash-dotted: $\sqrt{\pi}/32$. The crosses display the exact results using the full Fock state basis.}\label{mode4_30_sg}
\end{figure}

Next we extend our results to the four-mode case with a constant in time hopping parameter $J=1$. In Fig.~(\ref{mode4_30_sg}) results are shown for the initial state $|30,-\sqrt{0.7},\sqrt{0.3},0,0\rangle$. Although it has less photons than the in two-mode case above,
the spacing$\sqrt{\pi}/4$ which performs well in Fig.~(\ref{mode2_50_sg}) does not give rise to converged results with 169 basis functions, and this kind of deviation for $\sqrt{\pi}/4$ also occurs for other states with larger numbers of photons or other initial state parameters (not shown).
Thus the number of modes seems to be decisive, when choosing the optimal grid spacing.

\begin{figure}[h]
  \centering
  \includegraphics[width=3in]{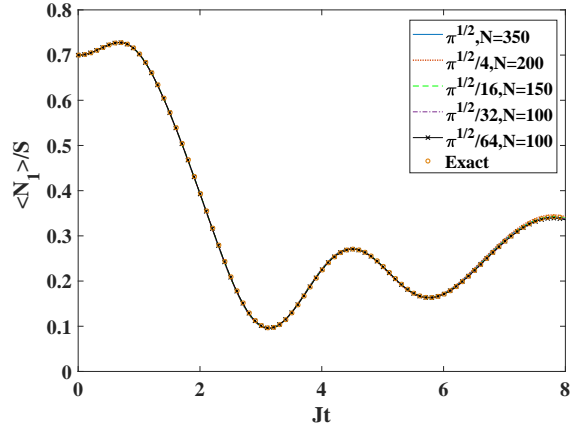}\\
  \caption{Convergence of the population dynamics in the four-mode case with 30 photons for different spacings and basis size $N$. The on-site interaction energy $U=0.1J$. The different curves (solid: $\sqrt{\pi}$, dotted: $\sqrt{\pi}/4$, dash: $\sqrt{\pi}/16$, dash-dotted: $\sqrt{\pi}/32$, star: $\sqrt{\pi}/64$,
  circles: exact Fock results) coincide to within line thickness for most of the
  time interval shown.}\label{mode4_30_cp_sg}
\end{figure}

In Fig.~(\ref{mode4_30_cp_sg}), we present a study of the number of basis functions which is needed for getting converged results
by optimizing the underlying grid spacing. This time, however, in order for faster convergence, we go back to the case of random grids employed in the main text. All results initially coincide to within line thickness and only for longer times the small spacing turns out to be advantageous.
Although small spacings make the calculation generally more efficient, decreasing the spacing indefinitely is not an option for improvement and we found an optimum value of $\sqrt{\pi}/32$  in the present case. The even smaller spacing $\sqrt{\pi}/64$ will not lead to a further promotion
compared with $\sqrt{\pi}/32$ (the results are even slightly worse). Our final conclusion is that
large numbers of photons and modes will increase the dimension of Hilbert space dramatically (see Eq. (\ref{eq:choose})) and make the dynamical process more complicated. As we have shown here,
constructing complex grids with optimal spacings (smaller than $\sqrt{\pi}$) can help speeding up the convergence of numerical calculations, and for more complicated systems smaller spacing is needed for feasibility of the calculation, but there is no gain in decreasing the spacing indefinitely.

\end{appendix}

\newpage

%

\end{document}